\documentstyle[prl,aps,multicol,epsf]{revtex}
\begin{document}
\draft
\widetext
\title{Ferromagnetic transition in a 
double-exchange system containing impurities in  the Dynamical Mean 
Field Approximation}
\author{Mark Auslender$^1$ and Eugene Kogan$^{2}$}
\address{$^1$ Department of Electrical and Computer Engineering,
Ben-Gurion University of the Negev,
P.O.B. 653, Beer-Sheva, 84105 Israel\\
$^2$ Jack and Pearl Resnick Institute 
of Advanced Technology,
Department of Physics, Bar-Ilan University, Ramat-Gan 52900, 
Israel}
\date{\today}
\maketitle
\begin{abstract}
\leftskip 54.8pt
\rightskip 54.8pt
We formulate the Dynamical Mean 
Field Approximation equations for 
the double-exchange system with  quenched disorder for arbitrary relation
between Hund exchange coupling and electron band width. 
Close to the
ferromagnetic-paramagnetic transition point the DMFA equations can be
reduced to the ordinary mean field equation of  Curie-Weiss type.
We  solve the
equation to find the  transition temperature and present the 
magnetic phase diagram of the system. 
\end{abstract}
\begin{multicols}{2}
\narrowtext

\section{Introduction}

The Dynamical Mean Field Approximation (DMFA) (see
\cite{DMFA} and references therein) is a widely used formalism for 
many body systems. In many cases it well
takes into account non-trivial effects of strong correlations and thermal
disorder.
In particular, for  the double-exchange (DE) 
model \cite{zener,anderson}, where the 
exchange interaction between the core
spins is mediated by mobile carriers, application of the DMFA 
to the description of
ferromagnetic transition 
(see Ref. \cite{furukawa2}
and references therein) resolved a long standing problem of the energy scale for
the Curie temperature $T_{\rm c}$ of manganites.
However, this application  was restricted to the case where  only 
thermal magnetic  disorder is
present in the model. Thus, chemical
disorder introduced by doping impurities, which is generic for 
the manganites and many other DE systems,
was ignored. 

It was shown however, that the
concurrent action of the static and magnetic disorder  is crucial for the
the description of the quasi-particle and transport properties of 
the DE  systems \cite{dong,auskog,gorkov}. 
Therefore the  inclusion of static disorder 
into the DMFA scheme is actual. The present note is devoted to this task.

\section{Hamiltonian and DMFA equations}

Consider  the DE model with random on-site energies. The Hamiltonian  of
the model  is 
\begin{eqnarray}
\label{HamDXM}  
H = \sum_{nn'\alpha} t_{n-n'} c_{n\alpha}^{\dagger} c_{n\alpha}+
\sum_{n\alpha}V_n c_{n\alpha}^{\dagger} c_{n'\alpha}\nonumber\\ 
-J \sum_{n\alpha\beta} {\bf m}_n\cdot 
{\bf \sigma}_{\alpha\beta}c_{n\alpha}^{\dagger} c_{n\beta},
\end{eqnarray}
where $t_{n-n'}$ is the electron hopping, $V_n$ is the random
on-site energy, $J$ is the effective 
exchange 
coupling between a core spin and a conduction electron,
$\hat{\bf \sigma}$ is the vector of the Pauli matrices, and $\alpha,\beta$ are
spin indices. 
We express the localized (classical) spin  by
 $ {\bf m}_n = (m_n{}^x, m_n{}^y, m_n{}^z)$ 
with the normalization $|{\bf m}|^2 = 1$.

In a single electron representation the Hamiltonian 
 can be presented as
\begin{equation}
\label{generic}
H_{nn'}=H^0_{n-n'}+\left(V_n-J {\bf m}_n\cdot {\bf \sigma}\right)\delta_{nn'};
\end{equation}
the first  is  translationaly invariant, the second describes
quenched disorder, and the third - annealed disorder.

The DMFA, as applied to the problem under consideration,  is based on two assumptions. The first assumption is 
that the averaged, 
with respect to   random orientation of localized spins and
random on-site energy $V$,  locator 
\begin{eqnarray}
\hat{G}_{\rm loc}(z)= 
\left\langle\hat{G}_{nn}(z)\right\rangle_{{\bf m},V},
\end{eqnarray}
where
\begin{eqnarray}
\label{green}
\hat{G}(z)=(z-H)^{-1}, 
\end{eqnarray}
can be expressed through the  the local self-energy $\hat{\Sigma}$ by the
equation
\begin{eqnarray}
\label{local}
\hat{G}_{\rm loc}(z) =g_0\left(z - \hat{\Sigma}(z)\right),
\end{eqnarray}
where
\begin{eqnarray}
\label{g}
g_0(z) =\frac{1}{N}\sum_{\bf k}\left(z-H^0_{\bf k}\right)^{-1} 
\end{eqnarray}
is the bare (in the
absence of the disorder and  exchange interaction) locator. Thus introduced
self-energy satisfies equation
\begin{eqnarray}
\hat{G}_{\rm loc}(z)=\left\langle \frac{1}
{\hat{G}_{\rm loc}^{-1}(z)+\hat{\Sigma} (z)- 
V_n+J{\bf m}\cdot\hat{\bf \sigma}}\right\rangle_{{\bf m},V}.
\label{cpa}
\end{eqnarray}
The system of equations (\ref{local}) and  (\ref{cpa}) is very much similar to 
the well known CPA equations (see \cite{ziman} and references therein), as
generalized to the case when  the quantities  $\hat{G},\hat{\Sigma}$
and $\hat{g}$ are $2\times 2$ matrices in spin space \cite{kubo}.
The system of equations however, is not yet closed. The averaging with respect
to annealed disorder 
is principally different from the averaging with respect
to quenched disorder.

The second assumption of the DMFA is the prescription for the
determining, in our case, the probability 
of a spin configuration self-consistently with the solutions of
the Eqs. (\ref{local}) and (\ref{cpa}).
To formulate the DMFA equation for this probability, taking into account both
kinds of the disorder, 
let us start from the general formula for
the partition  function 
\begin{equation}
{\cal Z}_{V_n}=\int \exp \left( -{\rm Tr}\sum_s\log \hat{G}(z_s) \right)
\prod_{n}d{\bf m}_{n},
\label{grandz}
\end{equation}
where  $z_s = i\omega_s + \mu$;  $\omega_s$ is  the Matsubara frequency 
and $\mu$ is the chemical potential. 
The averaging over $\left\{{\bf m}_n \right\}$ is given by
\begin{eqnarray}
\left\langle \Phi \right\rangle _{\bf m}=\frac{1}{{\cal Z}_V}\int
\exp \left(-{\rm Tr}\sum_s\log \hat{G}(z_s)\right)
\Phi({\bf m})\prod_{n}d{\bf m}_{n}.
\label{funav}
\end{eqnarray}
All observables, in particular thermodynamic potential $\Omega$, should 
additionally be
averaged over the realizations of the quenched disorder; in particular
\begin{equation}
\Omega=-\frac{1}{\beta }\left\langle \log {\cal Z}_V
\right\rangle _{{\bf m},V}.  
\label{freenergy}
\end{equation}
The DMFA approximates the multi-spin probability 
${\cal Z}_V^{-1}\exp \left(-{\rm Tr}\log
\hat{G}\right)$ as a product of one-site probabilities in such a way, that
\begin{equation}
\frac{\delta \Omega}{\delta \hat{G}_{\rm loc}}=0.
\end{equation} 
The result for the one-site probability reads (for details of the
calculation see Ref. \cite{ak2}):
\begin{eqnarray}
\label{prob2}
P_{V_n}({\bf m})\propto \exp\left[-\beta\Delta\Omega_{{\bf m},V_n}\right],
\end{eqnarray}
where 
\begin{eqnarray}
\label{probability}
\Delta\Omega_{{\bf m},V_n}= -\frac{1}{\beta}\sum_s {\rm Tr} \log 
\left[1+\hat{G}_{\rm loc}(z_s)\right.\nonumber\\
\left.\left(
J{\bf m}\cdot\hat{\bf \sigma}- 
V_n+\hat{\Sigma} (z_s)\right)\right] e^{i\omega_s 0_{+}}.
\end{eqnarray}
is the change of the  thermodynamic potential 
of the electron
gas described by the Green's function $\hat G_{\rm loc}$ 
due to interaction with a single impurity \cite{doniach,chat}.

The right hand side of Eq. (\ref{prob2}),  
is a complicated non-linear functional of $P_V({\bf m})$.
However, 
 if we
are interested only in the transition
temperature $T_{c}$,  the problem can be reduced to a
traditional mean field (MF) equation.
In   linear  with
respect to magnetization $M$ approximation   Eq. (\ref{prob2}) takes the form
\begin{eqnarray}
P_{V_n}({\bf m})\propto  \exp\left( -\beta I_{V_n}{\bf M}\cdot{\bf m}\right).
\label{probability2}
\end{eqnarray}
Non-trivial solution of the MF equation 
\begin{equation}
{\bf M}=\int \left\langle P_{V_n}({\bf m})\right\rangle_V {\bf m}d{\bf m}.
\end{equation}
 can exist only for 
$T<T_{\rm c}$, where $T_c=\frac{1}{3}\left\langle I_{V_n}\right\rangle_V$.

\section{$T_{\rm c}$ for the semi-circular DOS}

The problem of finding $T_c$ is reduced to calculation of $g_V$ and $h_V$.
For simplicity consider the semi-circular (SC) bare density of states (DOS)
$N_0(\varepsilon)$,  the bandwidth being $2W$.
Then
\begin{eqnarray}
\label{gint}
g_0(z)=\int \frac{N_0(\varepsilon)d \varepsilon }{z - \varepsilon}
=\frac{2}{W}\left[\frac{z}{W}-
\sqrt{\left(\frac{z}{W}\right)^{2}-1}\right].
\end{eqnarray}
For this case
\begin{equation}
\label{sigma}
\hat{\Sigma}=z-2w\hat{G}_{\rm loc}-\hat{G}_{\rm loc}^{-1},
\end{equation}
where $w= W^2/8$.
Thus from Eqs. (\ref{local}) and (\ref{cpa}) we  obtain a single equation
 for $\hat{G}_{\rm loc}$
\begin{equation}
\label{sigma2}
\hat{G}_{\rm loc}(z)=\left\langle\frac{1}{z-2w\hat{G}_{\rm loc}(z)- V_n
+J{\bf m}\cdot\hat{\bf \sigma}}\right\rangle_{{\bf m},V},
\end{equation}
and Eq. (\ref{probability}) can be presented as
\begin{eqnarray}
\label{probability4}
\Delta\Omega_{{\bf m},V_n}= \frac{1}{\beta}\sum_s \log \det
\left[z_s-2w\hat{G}_{\rm loc}(z_s)\right.\nonumber\\
\left.- V_n
+J{\bf m}\cdot\hat{\bf \sigma}\right] e^{i\omega_s 0_{+}}.
\end{eqnarray}
In linear with respect to $M$ approximation
\begin{equation}
\hat{G}_{\rm loc}= g \hat I -
hJ{\bf M\cdot \hat\sigma},
\end{equation}
where $g$ is locator in paramagnetic phase, given by the equation
\begin{eqnarray}
\label{rq2}
g=
\frac{1}{2}\left[\left\langle\frac{1}{z-2wg-V_n-J}\right\rangle_{V}\right.
\nonumber\\
\left.+ \left\langle\frac{1}{z-2wg-V_n+J}\right\rangle_{V}\right],
\end{eqnarray}
and the quantity $h$  is given by the formula
\begin{eqnarray}
\label{hv}
h = \frac{\left\langle \Delta_{V_n}\right\rangle_V}
{1 -\frac{4J^2w}{3}\left\langle\Delta_{V_n}^2\right\rangle_V 
-2w\left\langle\Delta_{V_n}\right\rangle_V},
\end{eqnarray}
where  
\begin{equation}
\Delta_{V_n}(z_s)=\frac{1}{\left[z_s-2wg(z_s)-V_n\right]^2 -J^2}.
\end{equation}
Expanding Eq. (\ref{probability4}) we obtain
the effective exchange integral  is   
\begin{eqnarray}
I_{V_n}= \frac{4J^2w}{\beta}\sum_s h(z_s)\Delta_{V_n}(z_s).
\label{probability3}
\end{eqnarray}
If we transform the sum
over the imaginary Matsubara frequencies in the right-hand side of 
Eq. (\ref{probability3}) to integral over real energies $E$, we obtain for the
$T_{\rm c}$  
\begin{eqnarray}
\label{Theta}
T_{\rm c}=\frac{4J^2w}{3\pi}\int_{-\infty}^{\infty}f(E)
\mbox{Im}\left[h(E_+)\left\langle \Delta_{V_n}(E_+) \right\rangle_V\right]dE,	
\end{eqnarray}
where $f(E)$ is the Fermi function, and $E_+= E+i0$.
Eq. (\ref{Theta}), giving the Curie temperature as a function of the parameters
of the system, is the main result of the present work. 

It is worth analyzing the limiting cases of this equation.
In  $J\gg
W$ limit, shifting the energy by $J$, 
we obtain Eq. (\ref{Theta}) in the form  \cite{furukawa,ak}
\begin{eqnarray}
\label{theta}
T_{\rm c}=\frac{4w}{\pi}\int_{-\infty}^{\infty}f(E)
\mbox{Im}\left[\frac{ \left\langle g_{V_n}(E_+)\right\rangle_{V}^2}
        {3 - w\left\langle g_{V_n}(E_+)^2\right\rangle_{V}}\right]dE,	
\end{eqnarray}
where $g_{V_n}(E)=(E-w\left\langle g_{V_n}\right\rangle_{V}-V_n)^{-1}$.
In the Appendix we compare the 
$J\ll W$ limit of Eq. (\ref{Theta}) with the results of the
Ruderman-Kittel-Kasuya-Yosida (RKKY) theory \cite{vf}.

\section{Phase diagram}

Consider first the the phase diagram
(PD) of the system in case of no quenched disorder.
The PD is presented on Fig. 1.
\begin{figure}
\epsfxsize=3truein
\centerline{\epsffile{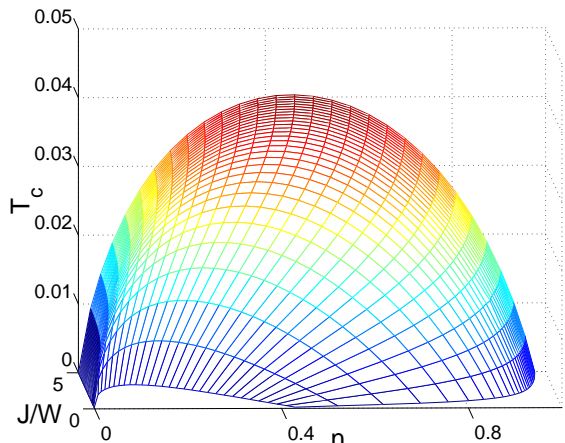}}
\label{Fig.1}
\caption{The phase diagram for the case of no quenched disorder
in the coordinates of
relative strength of the Hund exchange $J/W$ and electron concentration $n$.} 
\end{figure}
In the region, where Eq. (\ref{Theta}), gives negative value for the  $T_c$,
ferromagnetism is
precluded  at any temperature. 
From our consideration we can  say nothing about the nature of the 
non-ferromagnetic
phase (or phases), but we know from the theory of the RKKY interaction
\cite{mattis}, that for small $J/W$ (and no disorder), the ground state for the
intermediate electron concentration is
antiferromagnetic. One can say that  the situation with  finite Hund exchange is
equivalent in some sense to the situation with the infinite Hund exchange and
antiferromagnetic superexchange \cite{alonso}.

We consider the model of the disorder in which $V_n=V$ with the probability $x$,
and
$V_n=0$ with the probability $1-x$, thus $x$ being the concentration of
impurities. Solving equation for the locator for the case of strong quenched
disorder ($V/W=1$ and $x=.3$) we obtain the PD, which is
presented on Fig. 2.
\begin{figure}
\epsfxsize=3truein
\centerline{\epsffile{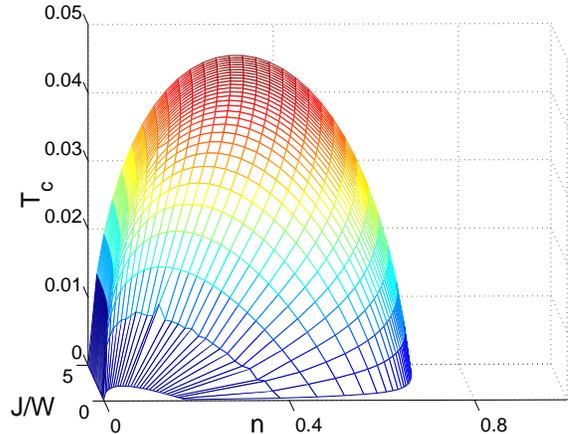}}
\label{Fig.2}
\caption{The phase diagram for $V/W=1$ and $x=.3$
in the coordinates of
relative strength of the Hund exchange $J/W$ and electron concentration $n$.} 
\end{figure}
It is interesting that ferromagnetism is now precluded in much larger region of
the $J/W-n$ plane.

This research was supported by the Israeli Science Foundation administered
by the Israel Academy of Sciences and Humanities. 

\appendix
\section{DMFA vs RKKY}

In $J\ll W$ limit from Eq. (\ref{Theta})
 after simple algebra we obtain
\begin{eqnarray}
\label{rkky}
T_{\rm c} = \frac{2J^2}{3}\int_{-\infty}^{\infty}f(E)
\left\{\frac{d\left\langle N_V(E)\right\rangle_{V}}{dE}\right.\nonumber\\
\left.-\frac{1}{\pi}\mbox{Im}
\left\langle  g_V(E_+)^2\right\rangle_{V}\right\}dE,
\end{eqnarray}
where  $N_V(E)=-(1/\pi)\mbox{Im}\;g_V(E_+)$ is the DOS (per one spin direction),
and
$g_V(E)$ is the solution of the equation 
\begin{eqnarray}
g_V(E)=\frac{1}{E-2w\left\langle g_V\right\rangle_{V}-V}.
\end{eqnarray}

In fact, in the weak exchange case we can 
expand Eq. (\ref{green})  with respect to exchange interaction. 
The second order term, which is
only  important for us, for the diagonal matrix element of the Green's function
 $G$ is
\begin{eqnarray}
\label{expansion}
\hat G^{(2)}_{ll}=\frac{J^2}{N}\sum_{l'nn'}\frac
{<ll'|nn'>}{(E-\epsilon_{l})^2(E-\epsilon_{l'})}({\bf m}_n\cdot {\bf
\hat\sigma})
({\bf m}_{n'}\cdot {\bf \hat\sigma}),
\end{eqnarray}
where 
\begin{equation}
<ll'|nn'>\equiv \psi^*_l(n) \psi_{l'}(n)\psi^*_{l'}(n')\psi_l(n'),
\end{equation}
and $\psi$ are the eigenfunctions of the non-magnetic part of the Hamiltonian:
$(H_0+V)\psi_l=\epsilon_l\psi_l$.
The thermodynamic potential of the electron subsystem 
 is given by the formula
\begin{equation}
\label{f2}
\Omega^{(2)}=\frac{1}{\pi\beta}\int_{-\infty}^{+\infty}\ln
\left[1+e^{-\beta(E-\mu)}\right]{\rm Im}\sum_l{\rm Tr}\;\hat
G^{(2)}_{ll}(E_+)dE.
\end{equation}
Calculating ${\rm Tr}$  with
respect to spin indices
and integrating by parts we obtain  
\begin{equation}
\label{h}
\Omega^{(2)}=-\sum_{n\neq n'}I_{nn'}{\bf m}_n\cdot {\bf m}_{n'},
\end{equation}
where
\begin{eqnarray}
\label{rs}
I_{nn'}=\frac{J^2}{\pi}\int_{-\infty}^{+\infty}f(E)
{\rm Im}\left[
\sum_{ll'}\frac{<ll'|nn'>}
{(E_+-\epsilon_{l})(E_+-\epsilon_{l'})}\right]dE.
\end{eqnarray}
The  exchange integral, after fulfilling in Eq. (\ref{rs}) 
integration with respect to
$dE$, can be presented as
\begin{equation}
\label{rs2}
I_{nn'}=J^2\sum_{ll'}\frac{f(\epsilon_{l'})-f(\epsilon_{l})}
{\epsilon_{l}-\epsilon_{l'}}<ll'|nn'>.
\end{equation}

To see the connection between the RKKY approximation and  Eq. (\ref{rkky}),
let us make in Eq. (\ref{h}) a mean field approximation
${\bf m}_n\cdot {\bf m}_{n'}=M^2$. Thus obtained potential can be used
to construct the Landau
functional  of the system, which gives \cite{ak}: 
\begin{equation}
T_c=-\frac{2}{3}\frac{\Omega^{(2)}}{NM^2}.
\end{equation}
Finally using the formula
\begin{equation}
\label{participation}
\sum_{n\neq n'}<ll'|nn'>
=\delta_{ll'}-\sum_n|\psi_l(n)|^2|\psi_{l'}(n)|^2,
\end{equation}
we obtain 
\begin{eqnarray}
\label{rkky2}
T_{\rm c} = \frac{2J^2}{3}\int_{-\infty}^{\infty}f(E)
\left\{\frac{dN^{(0)}(E)}{dE}\right.\nonumber\\
\left.-\frac{1}{\pi N}\mbox{Im}
\sum_n \left[G_{nn}^{(0)}(E_+)\right]^2\right\}dE,
\end{eqnarray}
where 
\begin{equation}
G^{(0)}_{nn}(E)=\sum_l\frac{|\psi_l(n)|^2}{E-\epsilon_l},
\end{equation}
and $N^{(0)}(E)=-(1/\pi N)\mbox{Im}\sum_nG^{(0)}_{nn}$.

Eqs. (\ref{rkky}) and (\ref{rkky2}) look very much alike. The only difference
between them is  ensemble averaging in Eq. (\ref{rkky}) vs site
averaging in Eq. (\ref{rkky2}). The DOS is self-averaging, that is
$N^{(0)}(E)=\left\langle N_V(E)\right\rangle_{V}$, because it involves the
locator
itself \cite{pastur}. Moreover, it is known that the CPA results for the DOS are
reasonable even in the case of strong disorder. So for low electron
concentration, when only the first term in Eq. (\ref{rkky})
(or Eq. (\ref{rkky2})) is important, the equations are equivalent. 
For higher electron concentration
the term with the square of the 
locator decreases the $T_c$.
According to
the RKKY theory, the Curie temperature goes through zero at approximately 
$n=.25$ for the three principal cubic lattices \cite{mattis} (for the case of
no quenched
disorder). Eq. (\ref{rkky}) for this case gives critical concentration $n=.4$.
This comparison allows us to estimate the degree of
agreement between the results of the DMFA and RKKY theory for  the case considered. 
In the opposite case
of very strong quenched disorder, the difference between the square of the
locator in  Eqs. (\ref{rkky}) and 
(\ref{rkky2}) becomes even larger, due to
the effects of localization, which are absent in the CPA. But the influence of
the localization on the destruction of ferromagnetism in the DE model demands
additional consideration.

\end{multicols}
\end{document}